\documentclass[12pt]{article}
\usepackage{amssymb,amsmath,epsfig}
\makeatletter

\@addtoreset{equation}{section}
\def\section{\@startsection {section}{1}{\z@}{-2.5ex plus -1ex minus
 -.2ex}{1.3ex plus .2ex}{\large\bf}}
\def\subsection{\@startsection{subsection}{2}{\z@}{-2.25ex plus%
 -1ex minus -.2ex}{0.5ex plus .2ex}{\bf}}

\advance \voffset by -0.8in
\advance \hoffset by -0.6in
\newcommand{\RR}{\mathbb{R}}

\def\ba {\mathbf a}

\def\bx{\mathbf x}

\newcommand{\inv} {{-1}}

\begin{document}
\parskip 6pt
\parindent 0pt
%\begin{flushright}
%HWM-03-2\\
%EMPG-03-02\\
%gr-qc/yymmnnn
%\end{flushright}

\begin{center}
\baselineskip 24 pt
{\Large \bf Spacetime geometry in (2+1)-gravity via measurements with returning lightrays  }

\baselineskip 18pt

\vspace{1cm}
{\large C.~Meusburger\footnote{\tt  catherine.meusburger@uni-hamburg.de}

Department Mathematik\\
Universit\"at Hamburg\\
Bundestra\ss e 55, D-20146 Hamburg, Germany} \\

\vspace{0.5cm}

{12 January 2010}

\end{center}

\begin{abstract}
We consider an observer in a (2+1)-spacetime without matter and cosmological constant who measures spacetime geometry by emitting lightrays which return to him at a later time. We investigate several quantities associated with such lightrays: the return time, the 
 directions into which light needs to be emitted to return and the frequency shift between the lightray at its emission and its return. 
We derive explicit expressions for these quantities as functions on the reduced phase space and  show how they allow the observer to reconstruct the full geometry of the spacetime in finite eigentime. 
We comment on conceptual issues. In particular, we clarify the relation between these quantities and Dirac observables and show that  Wilson loops arise naturally in these quantities.
\end{abstract}

\section{Introduction}

 Gravity in (2+1) dimensions plays an important role as a toy model for quantum gravity. It
 allows one to investigate
  fundamental conceptual questions of quantum gravity in a simplified theory amenable to quantisation.
However, these efforts have been hindered by difficulties in extracting meaningful physics from the theory.

Although holonomy variables and Wilson loops provide an explicit parametrisation of the phase space of the theory and serve as a starting point for quantisation - for a discussion of their role in (2+1)-loop quantum gravity and combinatorial quantisation  see \cite{loopcs} - it is difficult to relate them to physically meaningful quantities that could be measured by an observer. 
Except for particularly simple cases such as the torus universe \cite{Carlipbook, gua}, it has remained unclear how quantities that could be measured by observers such as lengths, times or angles could be expressed as functions on the physical phase space.
 Moreover, it is a priori not clear which physically meaningful measurements an observer could perform in an empty spacetime without local degrees of freedom and how such measurements would allow him to distinguish different spacetimes.

 In this paper, we show how this gap can be closed. We define several realistic measurements   that could be performed by an observer in a (2+1)-spacetime without matter. Specifically,
we consider an observer who emits lightrays which return to him at a later time. We investigate several measurements associated with such returning lightrays:
the return time, the directions into which light needs to be emitted to return and the frequency shift between the emitted and returning lightray. We derive explicit expressions
for these measurements as functions on the physical phase space and show that they allow the observer to reconstruct the full geometry of the spacetime in finite eigentime.

\section{Flat Lorentzian (2+1)-spacetimes}

\paragraph{Vacuum spacetimes  via the quotient construction} We consider maximally globally hyperbolic solutions of the three-dimensional Einstein equations without matter and cosmological constant. We restrict attention to spacetimes of topology $M\approx \mathbb R\times S_g$, where $S_g$ is a Riemann surface of genus $g\geq 2$.
These spacetimes have been classified by Mess \cite{mess}, for a more recent and accessible discussion see \cite{npm,bb, bg}. It is shown there
that their universal cover is an open, future complete domain $D\subset\mathbb M^3$ in Minkowski space, whose boundary $\partial D$ corresponds to the initial singularity of the spacetime. The domain is foliated by surfaces $D_T$ of constant cosmological time (CCT surfaces), i.e.~by surfaces of constant geodesic distance from the initial singularity. 

The fundamental group $\pi_1(M)\cong\pi_1(S_g)$ acts on the domain via a group homomorphism $h: \pi_1(M)\rightarrow P_3$ into the three-dimensional (proper orthochronous) Poincar\'e group $P_3=\text{Isom}(\mathbb M^3)$. Its  Lorentzian component
$h_L: \pi_1(M)\rightarrow SO^+(2,1)\subset P_3$ gives rise to
 a cocompact Fuchsian group $\Gamma$ of genus $g$, i.e. a discrete subgroup of the three-dimensional Lorentz group $SO^+(2,1)$ with $2g$ generators and a single defining relation
\begin{equation}
\Gamma=<a_1,b_1,...,a_g,b_g\;|\; b_g a_g^\inv b_g^\inv a_g\cdots b_1 a_1^\inv b_1^\inv a_1=1>\subset SO(2,1)^+.
\end{equation}
The action of $\pi_1(M)$ on the domain is free, properly discontinuous and preserves each surface of constant cosmological time. 
The  spacetime  $M\approx\mathbb R\times S_g$ is obtained as the quotient of the regular domain by this group action. It inherits a metric $g_M=-dT^2+g_{M_T}$ induced by the metric on Minkowski space and a foliation by CCT surfaces $M_T=D_T/h(\pi_1(M))$.  

\paragraph{Conformally static spacetimes}
The simplest spacetimes obtained in this way are the conformally static spacetimes, for which the domain is the interior of a future lightcone based at a point $\mathbf p \in \mathbb M^3$. The CCT surfaces foliating the domain are hyperboloids 
\begin{align}
D_T=\{ \mathbf y\in\mathbb M^3\:|\; (\mathbf y-\mathbf p)^2=-(y_0-p_0)^2+(y_1-p_1)^2+(y_2-p_2)^2=T^2\}\cong T\cdot \mathbb H^2.\nonumber
\end{align}
As the action of $\pi_1(M)$ preserves these hyperboloids, its translational component is trivial up to conjugation
 $h(\lambda)=(1, \mathbf p)\cdot(v_\lambda, 0)\cdot(1, -\mathbf p)$.  The action of its Lorentzian component 
on the hyperboloids coincides with the canonical action of the cocompact Fuchsian group $\Gamma$ on hyperbolic space $\mathbb H^2$.
The quotient of a cosmological time surface $D_T$ by this group action is therefore a Riemann surface rescaled by the cosmological time: $M_T=D_T/h(\pi_1(M))=T\cdot \mathbb H^2/\Gamma$. The metric of the quotient spacetime takes the form $g=-dT^2+T^2g_\Sigma$. The geometry of the CCT surfaces  changes with time only by a rescaling,  and the  quotient spacetimes are conformally static.

\begin{figure}
(a)  \includegraphics[height=.2\textheight]{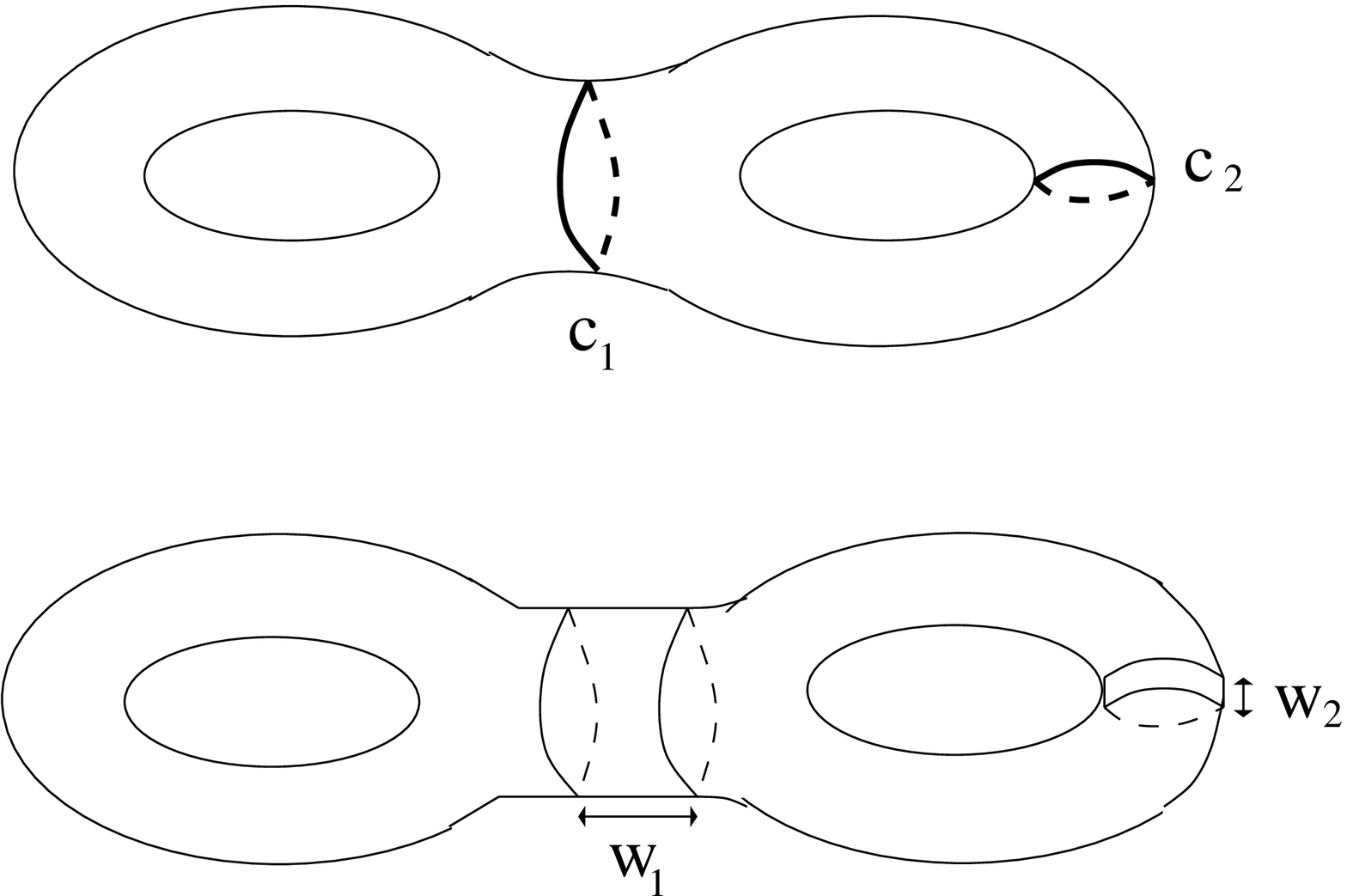}
(b)\includegraphics[height=.2\textheight]{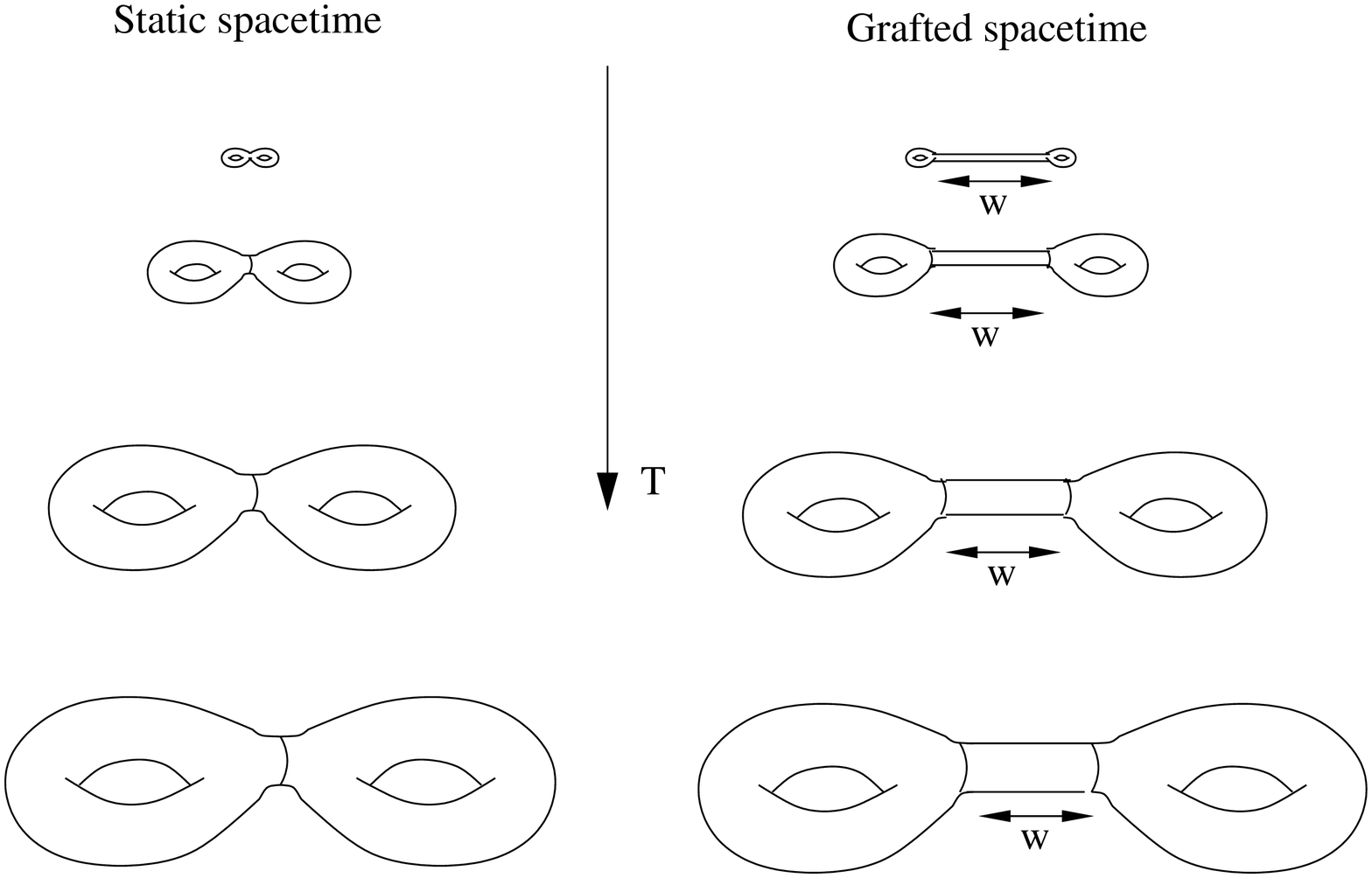}
  \caption{(a) Grafting on a Riemann surface, (b) Evolution of conformally static and grafted spacetimes with the cosmological time $T$.
  }
  \label{graft}
\end{figure}

\paragraph{Evolving spacetimes via grafting} For general spacetimes,  the regular domain and the group homomorphism $\pi_1(S_g)\rightarrow P_3$ take a more complicated form.  They are obtained from the conformally static spacetimes via the grafting construction. The ingredients in this  construction are a cocompact Fuchsian group $\Gamma$ of genus $g$ and a measured geodesic lamination on the associated Riemann surface $\Sigma=\mathbb H^2/\Gamma.$ We sketch the construction for the case where the measured geodesic lamination is a weighted multicurve, a set of closed non-intersecting geodesics $g_i$ on $\Sigma$, each associated with a weight $w_i>0$.
Schematically, grafting on a Riemann surface amounts to cutting the surface along each geodesic 
in the multicurve and inserting a hyperbolic cylinder whose width is given by the  weight as shown in figure \ref{graft} (a). The grafted (2+1)-spacetime is constructed by  grafting  all CCT surfaces  of the conformally static spacetime simultaneously in such a way that the weights are the same for all CCT surfaces. 
The resulting  spacetimes have a non-trivial evolution with the cosmological time depicted schematically in figure \ref{graft} (b).  While the components outside of the strips are rescaled, the width of the strips remains constant. The effect of the grafting therefore becomes negligible in the limit $T\rightarrow \infty$, but is prominent near the initial singularity.

\paragraph{Classification of spacetimes via holonomies} It is shown in \cite{mess}  that maximally globally hyperbolic spacetimes $M\approx \mathbb R\times S_g$ are characterised uniquely by their group homomorphisms $h:\pi_1(M)\rightarrow P_3$. Two spacetimes are isometric if and only if their holonomies are related by global conjugation with $P_3$. Moreover, each group homomorphism $h:\pi_1(M)\rightarrow P_3$ whose Lorentzian component $h_L:\pi_1(M)\rightarrow SO^+(2,1)$  defines a Fuchsian group of genus $g$ gives rise to such a spacetime. This implies that the phase space of the theory is the quotient
$\mathcal P_{phys}=\text{Hom}_0(\pi_1(S), P_3)/P_3$, where the subscript $0$ denotes the restriction to group homomorphisms whose Lorentzian component is Fuchsian of genus $g$.

The images  of this group homomorphism are the Poincar\'e valued holonomies
along closed curves in $M$.
A complete set of diffeomorphism invariant observables is given by the Wilson loops, which are  conjugation invariant functions of the holonomies. In (2+1)-gravity, there are two canonical Wilson loop observables associated with each element $\lambda\in\pi_1(M)$, the {\em mass} observable $m_\lambda$ and the {\em spin} observable $s_\lambda$, which correspond to the two Casimir operators of the Poincar\'e algebra. In the parametrisation $h(\lambda)=(v_\lambda, \mathbf a_\lambda)$, $v=\exp( n^a_\lambda J_a)\in SO^+(2,1)$, $\mathbf a_\lambda\in\RR^3$ they  are given by  $m_\lambda^2=\mathbf n_\lambda^2$, $s_\lambda=\hat{\mathbf n} \cdot \mathbf a_\lambda$.
Note that $m_\lambda^2>0$ for all $\lambda\in\pi_1(M)$ since all elements of a genus $g$ Fuchsian group are hyperbolic.
The physical interpretation of these observables and their role as generators of symmetry transformations via the Poisson bracket is discussed in \cite{ich,ich3}.

\section{Measurements via returning lightrays}

%\paragraph{Measurements with lightrays}
Although the quotient description of vacuum spacetimes provides an explicit characterisation of these spacetimes in terms of holonomies, it remains unclear how the holonomies  are related to quantities with a clear physical interpretation that could be measured by an observer in the spacetime. To obtain such quantities and relate them to the holonomies, 
we consider an observer who probes the geometry of the spacetimes by emitting ``test lightrays". Some of these lightrays return to the observer, who can then perform several measurements associated with such returning lightrays. He can determine  the eigentime elapsed between the emission of the lightray and its return,  the directions into which the light needs to be emitted  to return and the angles between them as well as the relative frequencies of the lightray at its emission and return.

%\paragraph{Observers}
In the following, we derive explicit expressions for such measurements and relate them to the physical degrees of freedom of the theory. For this, we consider an observer in free fall in the spacetime. The worldline of such an observer is a future directed, timelike geodesic in the quotient spacetime. This corresponds to an equivalence class of timelike, future directed geodesics in the  universal cover $D\subset M$, which is obtained by taking one lift  of this geodesic and acting on it with the holonomies. The resulting  geodesics are given by
\begin{equation}
\label{geodpar}
g_\lambda=h(\lambda)g\:\qquad\;\;g(t)=t\bx +\bx_0, \; \mathbf x^2=-1, \; \bx_0\in D, \lambda\in\pi_1(M).
\end{equation}
The two vectors $\bx\in\mathbb H^2$, $\bx_0\in D$ characterise the observer uniquely and can be interpreted, respectively, as his velocity unit vector and his initial position. Note that the  time coordinate $t$ corresponds to the observer's eigentime. 

In the following, we will need  further parameters, which are  functions of the vectors $\bx,\bx_0$  and of the holonomies $h(\lambda)$. The first is a rapidity parameter $\rho_\lambda$
\begin{equation}
\label{rapid}
\cosh \rho_\lambda=-\bx\cdot v_\lambda\bx\qquad \rho_\lambda=d_{hyp}(\bx, v_\lambda\bx),
\end{equation}
which is the hyperbolic distance between the velocity vectors $\bx,v_\lambda\bx\in\mathbb H^2$ and  the length of the associated geodesic on the CCT surface $\Sigma=\mathbb H^2/\Gamma$ of the conformally static spacetimes. We also introduce three parameters $\sigma_\lambda, \tau_\lambda, \mu_\lambda\in\RR$ which encode the relative initial position of the geodesic $g$ and its image $g_\lambda$
\begin{equation}
\label{transl}
h(\lambda)g(0)-g(0)=v_\lambda\bx_0-\bx_0+\ba_\lambda=\sigma_\lambda(v_\lambda \bx-\bx)+\tau_\lambda v_\lambda\bx+\mu_\lambda\bx\wedge v_\lambda\bx.
\end{equation}
%\paragraph{Returning lightrays}
A returning lightrays is a lightlike geodesic in the quotient spacetime, which intersects the worldline of the observer twice. This corresponds to a lightlike geodesic in the universal cover which is emitted at eigentime $t$ at the geodesic $g$  and arrives at one of its images $g_\lambda$ at eigentime $t+\Delta t$. It is characterised uniquely by the condition
\begin{equation}
\label{cond}
\left(h(\lambda)g(t+\Delta t)-g(t)\right)^2=0.
\end{equation}
Returning lightrays are therefore in one-to-one correspondence with elements of $\pi_1(M)$ and hence with closed geodesics on the constant cosmological time surfaces.

\section{Results and physical interpretation}

\paragraph{Measurements by observers as functions of the holonomies}
Using the parametrisation introduced above together with condition \eqref{cond},  one obtains an explicit expression for the eigentime $\Delta t$
elapsed between the emission of the lightray and its return
\begin{equation}
\label{rett}
\Delta t(t,\bx,\bx_0, h(\lambda))=(t-\sigma_\lambda)\cosh \rho_\lambda -\tau_\lambda+\sinh\rho_\lambda\sqrt{(t+\sigma_\lambda)^2+\mu_\lambda^2}.
\end{equation}
The direction into which the lightray needs to be emitted in order to return is given by the projection of the lightlike vector $h(\lambda)g(t+\Delta t)-g(t)$ on the orthogonal complement $\bx^\bot$ of the observer's velocity vector. It is characterised by a direction unit vector
\begin{align}
\label{unvec}
&\hat{\mathbf p}_\lambda(t,\bx,\bx_0, h(\lambda))=\cos\phi \frac{ \Pi_{\bx^\bot}(v_\lambda \bx)}{|| \Pi_{\bx^\bot}(v_\lambda \bx)||}+ \sin\phi \frac{\bx \wedge v_\lambda\bx}{||\bx\wedge v_\lambda\bx||}\\
&\text{with}\quad\label{direm}
\tan \phi(t,\bx,\bx_0, h(\lambda))= \frac{\mu_\lambda}{\sinh\rho_\lambda\sqrt{(t+\sigma_\lambda)^2+\mu_\lambda^2}+\cosh\rho_\lambda(t+\sigma_\lambda)}.
\end{align}
The relative frequencies of the lightray at its emission and return are given by expressions analogous to the one for  the relativistic Doppler effect \cite{gua, icht}. This yields \cite{icht} 
\begin{equation}
\label{doppler}
f_r/f_e(t,\bx,\bx_0, h(\lambda))=\frac{\sqrt{(t+\sigma_\lambda)^2+\mu_\lambda^2}}{\sinh\rho_\lambda(t+\sigma_\lambda)+\cosh\rho_\lambda\sqrt{(t+\sigma_\lambda)^2\!+\!\mu_\lambda^2}}.
\end{equation}
Formulae \eqref{rett}, \eqref{unvec}, \eqref{direm} and \eqref{doppler} give explicit expressions of the return time, the direction of emission and the frequency shift in terms of the eigentime $t$ at the emission of the lightray, the vectors $\bx,\bx_0$ which characterise the observer, and of the holonomies $h(\lambda)$, which characterise the spacetime. The expressions are independent of the choice of the lift and invariant under Poincar\'e transformations which act simultaneously on the geodesic of the observer and on the holonomies by conjugation.

\paragraph{Conformally static spacetimes}
To  understand how these measurements encode the geometry of the spacetime, we consider an observer in a conformally static spacetime for whom the eigentime coincides with the cosmological time. For such an observer, we have $h(\lambda)g(0)-g(0)=0$, which implies via \eqref{transl} $\sigma_\lambda=\tau_\lambda=\mu_\lambda=0$.  The expressions for the return time, the directions and the frequency shift simplify to   
\begin{equation}
\label{static}
\Delta t /t=e^{\rho_\lambda}-1\qquad \hat{\mathbf p}_\lambda=\Pi_{\bx^\bot}(v_\lambda \bx)\;\quad(\phi=0)\qquad f_r/f_e=e^{-\rho_\lambda}.
\end{equation}
The return time is thus proportional to the emission time, the directions of emission are constant and the frequency shift is a constant redshift which grows with the exponential of the length of the associated geodesic. For a general observer in a general spacetimes, these values are approached in the limit $t\rightarrow \infty$. 
%\begin{equation}
%\label{limit}
%\lim_{t\rightarrow\infty} \frac{ \Delta t}{t}= e^{\rho_\lambda-1}\qquad
% \lim_{t\rightarrow\infty} \hat{\mathbf p}_\lambda(t) = \Pi_{\bx^\bot}(v_\lambda x)\;(\lim_{t\rightarrow\infty}\phi(t)=0)\qquad
%  \lim_{t\rightarrow\infty} \frac{f_r}{f_e}(t)=e^{-\rho_\lambda}.\nonumber
%\end{equation}
This reflects the fact that the effect of grafting becomes negligible when the cosmological time tends to infinity.

\begin{figure}
\begin{centering}
 (a) \includegraphics[height=.2\textheight]{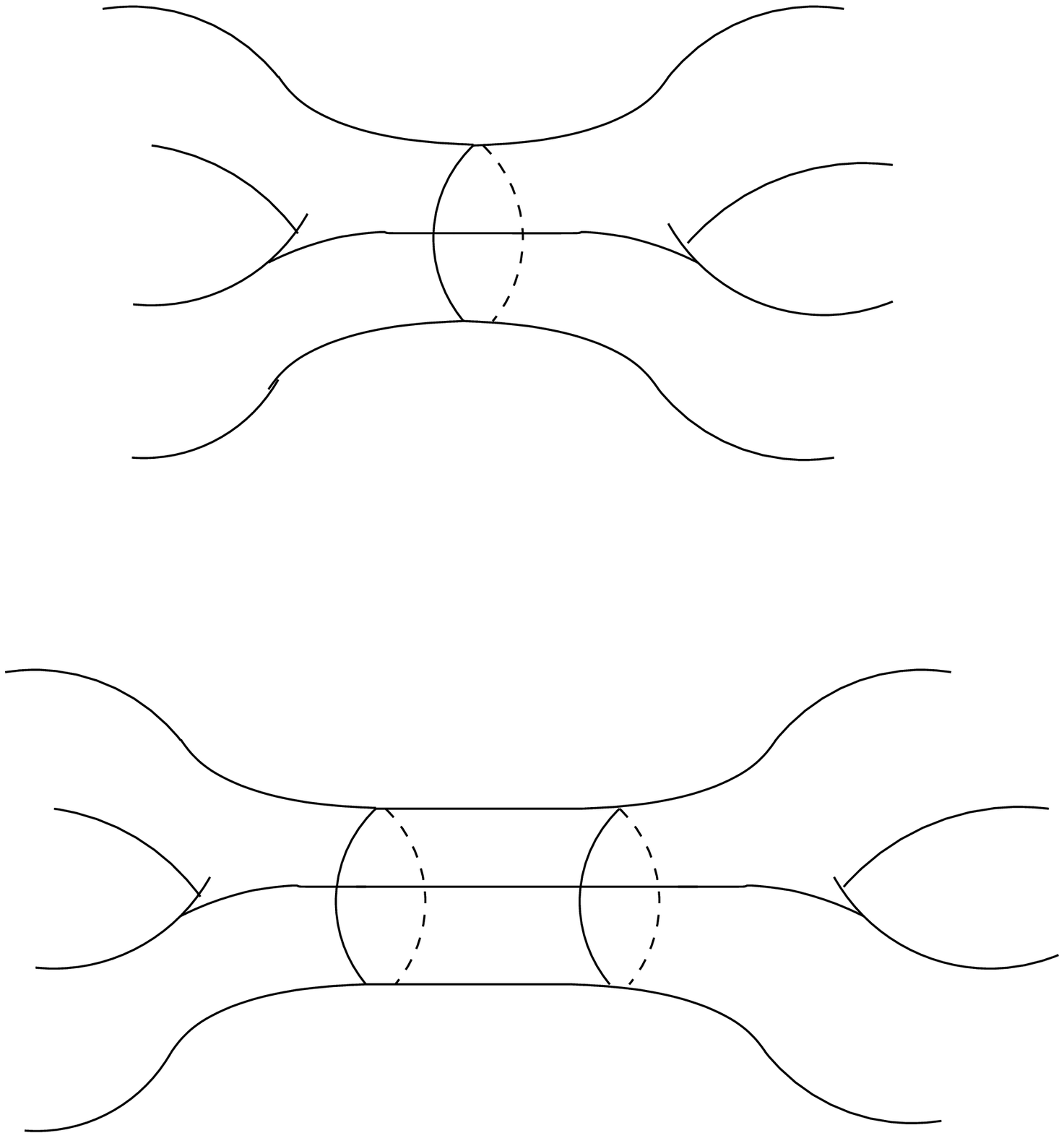}\hspace{.5cm}
 (b) \includegraphics[height=.2\textheight]{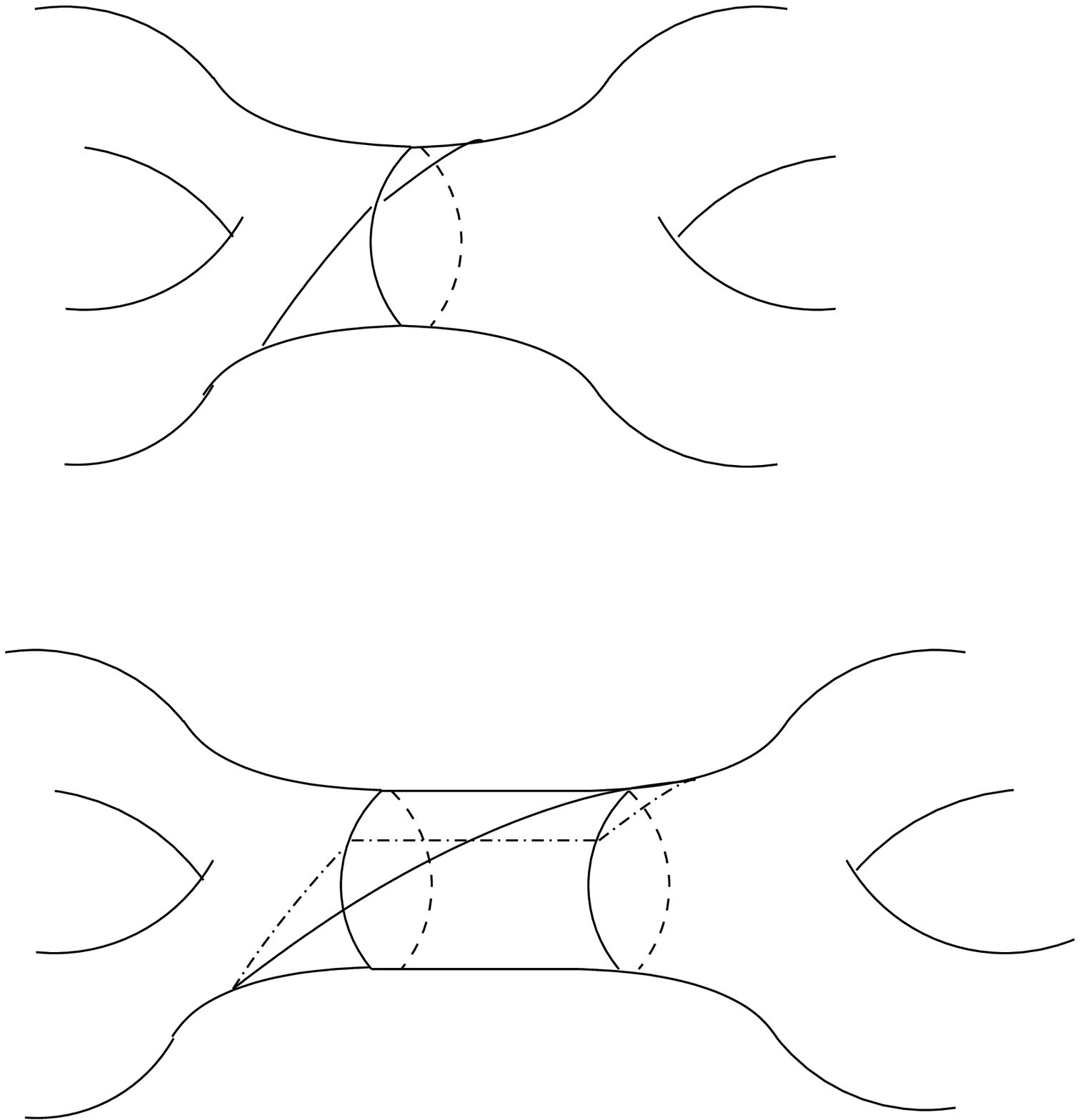}
  \caption{Effect of grafting on geodesics on a CCT surface. Case (a) corresponds to $\mu_\lambda=0$, case (b) to $\mu_\lambda\neq 0$
  }\end{centering}
  \label{graftgeo}
\end{figure}

\paragraph{Evolving spacetimes}To interpret the expressions for general spacetimes and observers, we note that it is the parameter $\mu_\lambda$, which is responsible for the nonlinearity in  equations \eqref{rett}, \eqref{direm} and \eqref{doppler},
while the parameters $\tau_\lambda$ and $\sigma_\lambda$  appear as a (lightray dependent)  constant in the expression for the return time and a (lightray dependent) shift in eigentime. It is shown in \cite{icht} that the parameter $\mu_\lambda$ is directly related to the grafting construction. If an evolving spacetime is constructed from a conformally static one via grafting, the parameter $\mu_\lambda$
for a given $\lambda\in\pi_1(M)$ vanishes if and only if the associated geodesics on a static CCT surface does not cross any grafting geodesics or crosses them orthogonally as shown in figure \ref{graftgeo} (a). In this case, the geodesic is not deflected at the grafting strip and its length increases by a constant.

If the geodesic associated with $\lambda\in\pi_1(M)$ crosses the grafting geodesic non-orthogonally as depicted in figure \ref{graftgeo} (b), we have $\mu_\lambda\neq 0$ and the grafting procedure causes a deflection of the geodesic as shown in figure \ref{graftgeo} (b). This deflection and the increase in length depend on  the emission time $t,$ since the width of the grafting strip remains constant while the rest of the spacetime is rescaled with the cosmological time $T$. This leads to a non-linear dependence of the return time on the emission time and to a time-dependence in the directions of emission and the frequency shift. The frequency shift is maximal if $\mu_\lambda$ vanishes but remains a red shift for all values of $\mu_\lambda$.

\section{Measurements vs observables}

An important issue in quantum gravity is the  relation between two notions of observables. The first is an observable as a quantity  measured by an observer. The second is that of a Dirac observable, i.e.~a function on the  reduced phase space
$\mathcal P_{phys}$.

While these two notions of observables  coincide in most physical systems, the situation is less clear in gravity.
This situation is reflected in formulas \eqref{rett}, \eqref{unvec}, \eqref{doppler}.  While it can be shown that these quantities are invariant under Poincar\'e transformations acting simultaneously on the holonomies and on the two vectors $\bx,\bx_0$, they are {\em not} invariant under conjugation of the holonomies alone. Although these quantities correspond to realistic measurements by observers, they are not functions on the physical phase space $\mathcal P_{phys}=Hom_0(\pi_1(S), P_3)/P_3$.

This is due to the fact that the  specification of an observer by two vectors $\bx$, $\bx_0$ is not physically meaningful  unless these vectors are related to a {\em physical}  quantity such as certain events in the spacetime.  To characterise an observer in a physically meaningful way, one needs to specify his worldline with respect to the geometry of the spacetime, i.e.~with respect to the holonomies $h(\lambda)$, $\lambda\in\pi_1(M)$.

The simplest way of implementing this is to introduce the notion of a ``comoving" observer with respect to a given holonomy $h(\lambda)$. This is an observer whose velocity unit vector $\bx$ lies in the plane stabilised by its Lorentz component $v_\lambda$, while the relative initial position vector \eqref{transl}
 is orthogonal to it. In the parametrisation $h(\lambda)=(v_\lambda, \mathbf a_\lambda)$, $v_\lambda=\exp( n^a_\lambda J_a)\in SO^+(2,1)$, $\mathbf a_\lambda\in\RR^3$ this amounts to
\begin{equation}
\mathbf n_\lambda\cdot \bx=0\qquad \mathbf n_\lambda\cdot(h(\lambda)g(0)-g(0))=\mathbf n_\lambda\cdot (v_\lambda g(0)-g(0)+\mathbf a_\lambda)=0.
\end{equation}
For such an observer, expressions \eqref{rett}, \eqref{unvec}, \eqref{direm}, \eqref{doppler} simplify to
\begin{align}
&\Delta t (t)=(\cosh m_\lambda-1)t+\sinh m_\lambda\sqrt{t^2 \sinh^2m_\lambda + s_\lambda^2}\\
&\hat{\mathbf p}_\lambda(t)=\cos\phi(t) \mathbf n_\lambda\wedge \bx+\sin\phi(t) \mathbf n_\lambda\qquad \tan\phi(t)=\frac{s_\lambda/\sinh m_\lambda}{t \cosh m_\lambda+\sqrt{t^2\sinh^2 m_\lambda+s_\lambda^2}}\nonumber\\
&f_r/f_e(t)=\frac{\sqrt{t^2\sinh^2 m_\lambda+ s_\lambda^2}}{t\sinh m_\lambda +\cosh m_\lambda\sqrt{ t^2\sinh^2 m_\lambda+s_\lambda^2}}\nonumber.
\end{align}
The expressions for the measurements thus take a particularly simple form 
for an observer who is  comoving with respect to a holonomy $h(\lambda)$. They
 are given by the two fundamental Wilson loop observables associated with $\lambda$. The mass observable $m_\lambda$ characterises them in the limit $t\rightarrow\infty$, while the spin $s_\lambda$ determines their behaviour near the initial singularity.

\section{ Reconstructing  spacetime geometry}

We will now demonstrate how these measurements allow the observer to determine the { full geometry} of the spacetime in  finite  eigentime. 
We focus on the case of conformally static spacetimes. The procedure for general spacetimes is similar but requires a finite number of additional measurements to determine the parameters which characterise their  evolution with the cosmological time.
As the phase space of the theory is $\mathcal P_{phys}=Hom_0(\pi_1(S), P_3)/P_3$, determining 
 the physical state of the universe amounts to measuring the holonomies $h(\lambda)$ for a set of generators of the fundamental group $\pi_1(M)$ up to conjugation. For conformally static spacetimes, in which their translational components are trivial, this amounts to determining a set of generators of the associated Fuchsian group $\Gamma$ given by their Lorentzian components.
 Such a set of generators can be obtained from its Dirichlet region.
This is the set of points in hyperbolic space $\mathbb H^2$ which are closer to a given point $\bx\in\mathbb H^2$ than to all its images under the action of $\Gamma$
\begin{equation}
D_\Gamma(\bx)=\{ \mathbf z\in\mathbb H^2 \:|\; d_{hyp}(\bx, \mathbf z)\leq d_{hyp}(v \bx,\mathbf z)\; \forall v\in \Gamma\}=\bigcap_{v\in\Gamma}\{\mathbf z\in \mathbb H^2\;|\; d_{hyp}(\bx, \mathbf z)\leq d_{hyp}(v \bx,\mathbf z)\}\nonumber.
\end{equation}
It is a geodesic arc polygon with $2k\geq 4g$ sides, which are identified pairwise by a set of generators of $\Gamma$. Determining a set of generators of the Fuchsian group is therefore equivalent to determining its Dirichlet region together with the information about  which pairs of sides are identified\footnote{I thank  R.~C.~Penner for pointing out that information about {\em the identification of  sides} is required in addition to the Dirichlet region itself. It can happen that two different Fuchsian groups have the same Dirichlet region but differ in the way they identify the sides of this region.}. 

To determine the Dirichlet region of the Fuchsian group, the observer emits light in all directions at a given time $t$. The lightrays will  return to him one by one. For each returning lightray, the observer records the return time $\Delta t(\lambda)$  and the direction from which the light returns to him. 
The return time $\Delta t(\lambda)$ determines the geodesic distance $\rho_\lambda=d_{hyp}(\bx, v_\lambda \bx)$ of $\bx, v_\lambda\bx \in\mathbb H^2$ via \eqref{static}. The  direction of return yields the tangent vector to the geodesic
 through  $\bx, v_\lambda\bx\in \mathbb H^2$ at $v_\lambda\bx$. Given the observer's velocity vector $\bx\in\mathbb H^2$, these two quantities allow him to uniquely determine the position of its image $v_\lambda\bx\in\mathbb H^2$. 
The observer can then construct the perpendicular bisector of the geodesic segment $[\bx, v_\lambda\bx]\in\mathbb H^2$ and the associated half plane $\{\mathbf z\in \mathbb H^2\;|\; d_{hyp}(\bx, \mathbf z)\leq d_{hyp}(v \bx,\mathbf z)\}$. Note that the observer does not need to know his velocity vector in order to do this. As the measurements are invariant under Lorentz transformations acting simultaneously on $\bx\in\mathbb H^2$ and on the holonomies by conjugation, a different choice of the vector $\bx$ leads to a global conjugation of the holonomies and thus to an equivalent Fuchsian group. 

After a finite number of returning lightrays,  the associated perpendicular bisectors form a geodesic arc polygon. For any image $v_\eta\bx$ whose geodesic distance from $\bx$ is greater than twice the maximal distance $d_{max}$ of $\mathbf x$ from the corners of this polygon, the perpendicular bisector of $[\bx, v_\eta\bx]$ cannot intersect the polygon. This implies that after a time interval $\Delta t=t(e^{2d_{max}}-1)$ the polygon cannot change anymore and coincides with the Dirichlet region. The observer can thus determine the Dirichlet region in a finite amount of eigentime.
Sending out a finite number of additional lightrays in the direction of the images $v_\lambda\bx$ associated with the boundary of the Dirichlet region and recording their return direction, the observer then obtains the identification of its sides and thus the holonomies $h(\lambda)$ for a set of generators of the fundamental group.

\section{Outlook and conclusions}

In this paper we defined several physically meaningful measurements that could be performed by an observer in a (2+1)-dimensional vacuum spacetime. We related them to the phase space variables used in the quantisation of the theory and  showed that they encode the full geometry of the spacetime. This provides a framework for the discussion of conceptual problems of (quantum) gravity.  

It would also be interesting to relate our results to the work of Budd and Loll \cite{lollbudd}, who obtain a quantisation of the theory without time variables, and to the results by Guadagnini \cite{gua} on the torus universe. 

It remains to investigate the quantities that correspond to these measurements in a quantum theory of Lorentzian (2+1)-gravity. This could be done either in a Hamiltonian quantisation formalism 
based on holonomy variables and Wilson loops or in a quantisation formalism that makes use of the fact that the phase space of the theory is a cotangent bundle on Teichm\"uller space such as  \cite{lollbudd}. Of particular interest in this respect would be the role of time and observers in the quantum theory and the question of mapping class group invariance.

%%%%%%%%%%%%%%%%%%%%%%%%%%%%%%%%%%%%%%%%%%%%%%%%
%% BACKMATTER
%%%%%%%%%%%%%%%%%%%%%%%%%%%%%%%%%%%%%%%%%%%%%%%%

\section*{Acknowledgments}
I thank the organisers of the XXV Max Born Symposium (Wroclaw, June 29 - July 3) where this work was presented as a talk. 
The research was partly undertaken  at  The University of Nottingham, partly at Hamburg University. The research at the University of Nottingham was supported by the Marie Curie Intra-European fellowship P IEF-GA-2008-220480. The research at Hamburg University is supported by the  Emmy Noether fellowship ME 3425/ 1-1 of the German Research Foundation (DFG).

%%%%%%%%%%%%%%%%%%%%%%%%%%%%%%%%%%%%%%%%%%%%%%%%
%% The bibliography can be prepared using the BibTeX program or
%% manually.
%%
%% The code below assumes that BibTeX is used.  If the bibliography is
%% produced without BibTeX comment out the following lines and see the
%% aipguide.pdf for further information.
%%
%% For your convenience a manually coded example is appended
%% after the \end{document}
%%%%%%%%%%%%%%%%%%%%%%%%%%%%%%%%%%%%%%%%%%%%%%%%

%%%%%%%%%%%%%%%%%%%%%%%%%%%%%%%%%%%%%%%%%%%%%%%%
%% You may have to change the BibTeX style below, depending on your
%% setup or preferences.
%%
%%
%% For The AIP proceedings layouts use either
%%%%%%%%%%%%%%%%%%%%%%%%%%%%%%%%%%%%%%%%%%%%

\bibliographystyle{aipproc}   % if natbib is available
%\bibliographystyle{aipprocl} % if natbib is missing

%%%%%%%%%%%%%%%%%%%%%%%%%%%%%%%%%%%%%%%%%%%
%% The following lines show an example how to produce a bibliography
%% without the help of the BibTeX program. This could be used instead
%% of the above.
%%%%%%%%%%%%%%%%%%%%%%%%%%%%%%%%%%%%%%%%%%%

\end{document}